# Public Cluster : parallel machine with multi-block approach

Z. Akbar[1*], Slamet[2], B.I. Ajinagoro[3], G.J. Ohara[3], I. Firmansyah[1], B. Hermanto[1], L.T. Handoko[1]

[1] *Group for Theoretical and Computational Physics, Research Center for Physics, Indonesian Institute of Sciences, Kompleks Puspiptek Serpong, Tangerang 15310, Indonesia*
[2] *Faculty of Computer Science, University of Indonesia, Kampus UI Depok, Depok 16424, Indonesia*
[3] *Faculty of Electronics Engineering, STT Telkom, Jl. Telekomunikasi, Dayeuh Kolot, Bandung 40257, Indonesia*

We introduce a new approach to enable an open and public parallel machine which is accessible for multi users with multi jobs belong to different blocks running at the same time. The concept is required especially for parallel machines which are dedicated for public use as implemented at the LIPI Public Cluster. We have deployed the simplest technique by running multi daemons of parallel processing engine with different configuration files specified for each user assigned to access the system, and also developed an integrated system to fully controll and monitor the whole system over web. A brief performance analysis is also given for Message Parsing Interface (MPI) engine. It is shown that the proposed approach is quite reliable and affect the whole performances only slightly.

## 1. Introduction

Along with the advances of scientific researches, especially in the field of basic natural sciences in the last decades, the needs on advanced computing is increasing exponentially. Such kind of advanced computings require higher specifcations on hardwares which leads then to astronomical cost to realize. One key solution for this problem is the parallel / cluster machine.

Nowadays, clustering (the low specs and low cost machines) becomes the mainstream to realize an advanced computing system comparable to or in most cases better than the conventional mainframe based system with significant reduced cost [1]. Generally the cluster is designed to perform a single (huge) computational task at certain period. This makes the cluster system is in general exclusive and not at the level of appropriate cost for most potential users, neither the young beginners nor the small research group, especially in the developing countries.

It is clear that the cluster is in that sense still costy, although there is certainly needs to educate the young generation to be the next users familiar with parallel programmings etc. This background motivates us to develop an open and free cluster system for public, namely the LIPI Public Cluster (LPC) [2,3,4].

Concerning those characteristics, the public cluster should be accessible and user-friendly for all users with various level of knowledges on parallel programming and also various ways of accessing the system in any platforms. This can be achieved by deploying the web-based interfaces in all aspects. Therefore, the main issues are then the security from anonymous users, avoiding the interferences among different tasks running on multi blocks of the nodes and the real-time monitoing and control over web for both administrators and users.

In this paper we present the simplest method to fill such requirements. In the subsequent section we first introduce the multi-block system, followed by our concept on public cluster. Finally we provide a brief performance test on the current system before coming to the conclusion.

## 2. Multi-block system

In order to run a particular task on several machines, we must implement an interface for the process management and communication among the nodes. In ur system we deploy the popular one, that is the Message-Passing Interface (MPI) [5]. MPI is widely used due to its portability to run MPI in any platforms through the message-passing protocol. Further, we have also implemeted the MPICH2 that is the upgraded version of the MPI developed by the Argonne National Laboratory [6]. InMPICH2 the process management and the communication is completely splitted off. So, the initial runtime environment contains several daemons called MPD's which each of them has a task to initiate the communication among the nodes before running the main programme. This mechanism is crucial since it enables us to distinguish the errors either in the process or communication.

Each MPD daemon is configurable through its configuration file or optional parameters in command line. Since in our system each user has their own configuration and daemon of MPD, MPICH suits well our needs on a cluster system with different parallel computation at the same time. Further, each user who is assigned to use the cluster with partiicular configuration and daemon is called as a *block*. Therefore, we define the multi-block as a case where several users with different blocks access the system at the same time.

MPD is a process management assembles daemons which execute the MPI modules. In order to form these daemons as a ring, there should be a node works as a master. This master node will boot the MPD in each node member. This flow is depicted in Fig. 1.

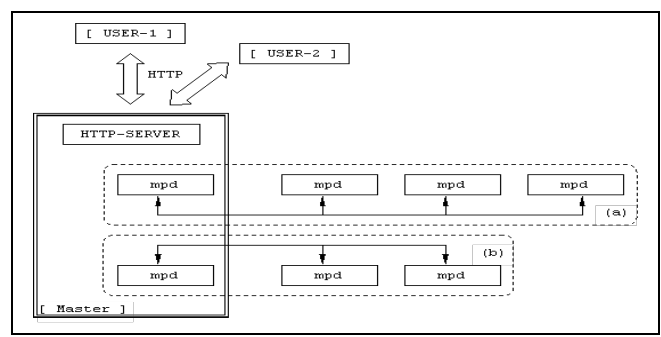

Fig. 1. The multi-block system at the LPC.

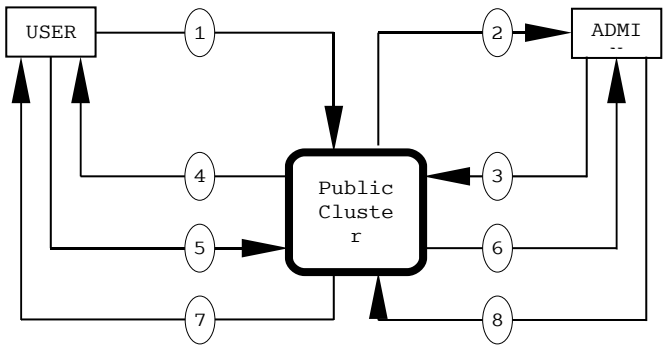

Fig. 2. The interaction flow of administrator and users at the LPC.

Fig. 1 shows logically the creation of two rings of MPD. The first block (a) contains three nodes in addition to one MPD in the master node. On the other hand, the second block (b) has two nodes and also one MPD in the master node. Each node is restricted to have only one running MPD, while the master node is allowed to run multiple MPD's as long as each MPD belongs to different user. Under this mechanism the interferences among the nodes could be completely avoided.

The mechanism to enable multi-block system requires several procedures :
- Each node, including the master node, communicates each other using passwordless access, for instance ssh with RSA-key. This can be accomplished easily without copying the authorized keys to all assigned nodes, for example by exporting the user's home directory through NFS.
- The node names which form the ring of MPD are located at the file `mpd.hosts` in the user's home directory. In the LPC, the node's names are assigned by the administrator and unchangeable by the users.
- The main configuration file for MPD, `.mpd.conf`, is located in the user's home directory. The important parameter is `MPD_SECRETWORD=<password>` which is crucial for security of the MPD's ring. The `<password>` should be the same for all nodes in the same ring.
- The master node behaves as dual-homed, that is it has multi network interfaces (NIC's) and IP addresses. Because we must split the NIC's off for the external and the internal accesses. The parameter to assign the appropriate NIC for the MPDdaemon is `--ifhn=<ip_address>`.
- It should be kept in mind that without any restriction, the number of processes running in a node is infinity and the tasks will be distributed in rond-robin way. This can be configured through the parameter `<node_name>:<process_number>`.

### 3. Public cluster

After discussing the multi-block system in the preceeding section, now let us consider the concept of public cluster.

First of all, let us recall the differences between the LPC and the other clusters in general :

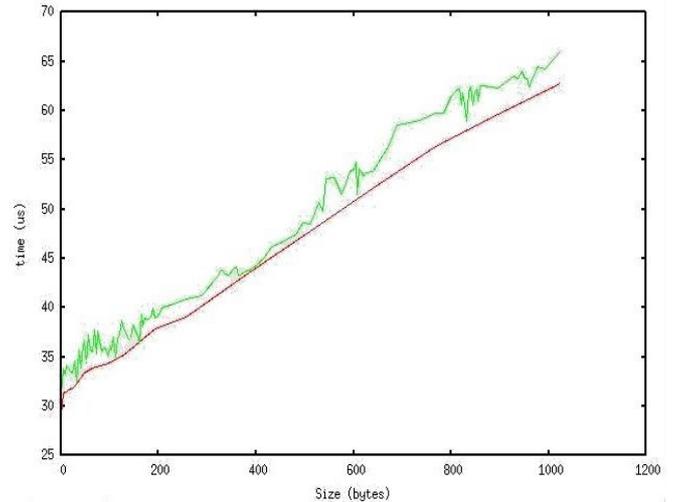

Fig. 3. Mesurement of bisection bandwith in single block (red line) and two blocks (green line) clusters.

- The users are all anonymous, i.e. everyone can register themself and then start using the cluster after getting approved and assigned with particular number of nodes by the administrator.
- The specification and the node number is fully determined by the administrator. So, the administrator has full control on the system, and then can shut unused nodes down if necessary to save the resources. This is crucial since the LPC consists of nodes with several specifications ranging from P4 with huge memories and spaces to the old 486 PC's. This means the administrator is able to assign an appropriate block of nodes concerning the level of computation will be perforned. In this sense, the distribution mechanism here is completely different with the known openPBS [7].
- The system is fully accessible, monitorable and controllable through the web interface by both administrator and users without having the ssh access. This would also improve the security concerning anonymous users accessing the system.

The work-flow of LPC is diagrammatically depicted in Fig. 2. The flow can be explained as below :
1. A new user does the registration by providing the personal data, the content of job will be performed and the number of nodes requested for the job.
2. The application is reviewed and verified by the administrator. The administrator will also assign the nodes provided for the approved user and the usage period.
3. Reconfirmation by user. If the user agrees with the provided nodes and the usage period, the administrator will switch the nodes on and activate all daemons.
4. The user should adjust their parallel programme to fit the provided nodes.
5. The user uploads all necessary programme and libraries if any. At this stage the programme can be executed immediately.
6. The administrator and automated system will monitor the usage of all running users.
7. Finished job can be downloaded by the user.

8. Once the usage period is over, the nodes are shutted down automatically.

## 4. Performance test

Now, we are ready to study the performance of the system. Our concern is of course how far the multi-block approach influences the whole performance ? This point is crucial, especially since all blocks (rings) rely on the same master node where it also acts as a node in each block.

The benchmark test is done by dividing the whole system to be two uncomparable blocks, i.e. there is a significant difference of node numbers between both blocks. As a benchmark tool, we use `mpptest` from the Argonne National Laboratory [8] and measure the bisection bandwith.

The benchmark is performed by running one block while the other is shutted down. Thereafter, both blocks are activated simultaneously. The result is shown in Fig. 3, where the red line is for single block, and the green line is for both blocks.

Form the figure, it is clear that running several blocks at the same master node is still reliable and affect the whole performances only slightly. Of course, this performance is not guaranteed for all kind of parallel programmings, but at least for certain objectives at the LPC this result proves and justifies the current architecture of LPC.

## 5. Conclusion

We have introduced an approach to enable multi-block system in a cluster dedicated for public use. Equipping the cluster with an integrated control and monitoring system, and also the interactive web interface would realize a low cost and easy-to-use or -maintain parallel machine for public. This would be a significant contribution to encourage and educate young generation on parallel programming which should be crucial for advanced researches in any areas in the future.

**Acknowledgment**

Slamet greatly appreciates the support from his colleagues at the Faculty of Computer Science, University of Indonesia. B.I. Ajinagoro and G.J. Ohara appreciate their colleagues at the Faculty of Engineering, STT Telkom. This work is financially supported by the Riset Kompetitif LIPI in fiscal year 2007.